\documentclass[12pt]{article}
\usepackage{amsmath,amssymb,amsthm,color}
\usepackage{graphicx}
\usepackage{cite}
\usepackage{epsfig}
\renewcommand{\baselinestretch}{2}

\begin{document}
%
\title{Feature-Rich Electronic Properties in Graphene Ripples \\}
\author{
\small Shih-Yang Lin$^{a}$, Shen-Lin Chang$^{a,*}$, Feng-Lin Shyu,$^{b,*}$ Jian-Ming Lu,$^{c}$ Ming-Fa Lin$^{a,*}$ $$\\
\small  $^a$Department of Physics, National Cheng Kung University, Tainan 701, Taiwan\\
\small  $^b$Department of Physics ROCMA, 830 Kaohsiung, Taiwan\\
\small  $^c$National Center for High-Performance Computing, Taiwan;  \\
\small      Department of Mechanical Engineering, Southern Taiwan University of Science and Technology, Taiwan \\
 }
\renewcommand{\baselinestretch}{1}
\maketitle

\renewcommand{\baselinestretch}{1.4}
\begin{abstract}

Graphene ripples possess peculiar essential properties owing to the strong chemical bonds, as an investigation using first principle calculations clearly revealed. Various charge distributions, bond lengths, energy bands, and densities of states strongly depend on the corrugation structures, ripple curvatures and periods. Armchair ripples belonging to a zero-gap semiconductor display split middle-energy states, while the zigzag ripples exhibit highly anisotropic energy bands, semi-metallic behavior implicated by the destruction of the Dirac cone, and the newly created critical points. Their density of states exhibit many low-lying prominent peaks and can explain the experimental measurements. There exist certain important similarities and differences between graphene ripples and carbon nanotubes.
\vskip 1.0 truecm
\par\noindent

\par\noindent  * Corresponding author. {~ Tel:~ +886-6-2757575-65272}\\~{{\it E-mail address}: mflin@mail.ncku.edu.tw (M.F. Lin)}
\end{abstract}

\pagebreak
\renewcommand{\baselinestretch}{2}
\newpage

{\bf 1. Introduction}
\vskip 0.3 truecm

Two-dimensional graphene has attracted considerable attentions in the fields of chemistry, material science and physics,\cite{novoselov2005two,geim2009graphene,lee2008measurement,stankovich2007synthesis,soldano2010production} and various studies have been conducted on its electronic,\cite{novoselov2005two,sprinkle2009first} transport,\cite{hwang2007carrier,sarma2011electronic} and magnetic\cite{zhang2006landau,checkelsky2008zero} properties. Graphene is viewed as a high-potential electronic material, because of its fascinating properties, extremely high mobility of charge carriers,\cite{bolotin2008ultrahigh,wang2009high} and the anomalous quantum Hall effect.\cite{zhang2005experimental,gusynin2005unconventional,du2009fractional} Its electronic properties can be tuned by the curvature of the structure,\cite{cortijo2007effects,lin2014magneto,chang2014geometric} the layer numbers,\cite{nilsson2006electronic} the stacking configurations,\cite{avetisyan2010stacking} the uniform or uniaxial mechanical strains,\cite{ni2008uniaxial,wong2012strain} and the application of external electric and magnetic fields.\cite{orlita2009graphite,chuang2013electric,wu2014combined} The natural non-uniformly corrugated structure of graphene can lead to certain peculiar changes in the electronic properties.\cite{fasolino2007intrinsic,meyer2007structure} The 2D graphene ripple, a uniform corrugated structure, has been the focus in surface science and material chemistry.\cite{meng2013strain,yan2013strain,capasso2014graphene} It is also expected to have a high potential applications in electronic devices and energy storages. This work investigates in detail for how the unique corrugated structure with various configurations induces the feature-rich electronic properties in graphene ripples (GRs).

Graphene-related materials with curved structures are successfully produced in laboratories, such as rippled,\cite{meng2013strain,yan2013strain,capasso2014graphene} bubbled,\cite{de2008periodically,bao2009controlled,georgiou2011graphene} scrolled,\cite{xie2009controlled,xia2010fabrication} and folded systems.\cite{liu2012folded,kim2011multiply} These systems can exhibit complex chemical bondings on the curved surface, and thus exhibit unique electronic spectra. The larger electronic orbitals on the outer side of the curved surface are suitable to adsorb atoms, for example, oxygen, hydrogen and lithium atoms.\cite{yoo2008large,yan2012high,tozzini2011reversible,hsiao2010preparation} Chemical doping on a curved structure might be more stable than on a planar one. Such curved environments are promising for potential application in future oxide semiconductors, hydrogen storage, and lithium batteries. We persue the full understanding of the curvature effects associated with a non-uniform charge distribution on the essential physical and chemical properties. These will play a critical role in the graphene-based application engineering.

A systematic study either theoretically or experimentally on graphene ripples' intricate electronic properties is needed. Previous theoretical research studies, mostly used the effective-mass model to deal with ideal GRs with an extremely rigid jump structure,\cite{guinea2008midgap,guinea2008gauge,guinea2009gauge,vozmediano2010gauge} but few have been based on first principle calculations with self-consistent optimal structures.\cite{wehling2008midgap} Instead, they mainly focused on is only on the effective magnetic fields or quasi-Landau levels. However, the longitudinal structures, the ratio of height and width, and periods were not taken into consideration, and the complete discussions on the band structures, density of states (DOS), and charge distributions are still incomplete. On the other hand, graphene ripples have recently been successfully grown on Rh(111) via the CVD process.\cite{meng2013strain,yan2013strain} Some of the results from scanning tunneling spectroscopy (STS) measurements exhibit low-lying peaks in the $dI/dV$-$V$ diagram.\cite{meng2013strain} In terms of the relationship between the peaks and the effective quantum numbers, it is not clear which ripple structure is the one responsible. The dispersions of energy bands which contribute to the quasi-Landau levels have not yet been identified.

In this paper, the geometric and electronic properties of the zigzag and armchair ripples are investigated in detail by first principles calculations. Among the two kinds of archiral structures, two basic types of electronic structures could be obtained, which could be further modulated by varying the ratio of height and width of ripples, and ripple periods. We seek to understand how many kinds of low-lying energy dispersions are exhibited, and how they are influenced by the geometric structures, including the presence or the absence of the Dirac cone, the curvature-induced two-dimensional highly anisotropic energy bands, and the association between the newly formed band-edge states and critical points. The curvature effects on low-energy band structures are studied thoroughly by analysis of the complex $\pi$ and $\sigma$ bondings. A comparison between GRs and carbon nanotubes is worth making since both are periodic systems. Moreover, how the change of rotational symmetry affects the middle-energy electronic properties is carefully examined. All the predicted band structures are further reflected in various feature-rich DOS. These characteristics in DOS could explain the previous experimental STS measurements;\cite{meng2013strain} in the meantime, we also anticipate some new features which will be further validated. The predictions of the anisotropic band dispersions could be further measured by angle-resolved photoemission spectroscopy (ARPES).\cite{ohta2007interlayer,ohta2006controlling} As expected, these rich fundamental features in graphene ripple structures can provide potential applications in energy materials or electronic devices.

\vskip 0.6 truecm
\par\noindent
{\bf 2. Methods }
\vskip 0.3 truecm

Our first-principles calculations are based on the density functional theory (DFT) implemented by the Vienna \emph{ab initio} simulation package (VASP).\cite{kresse1996efficient} The generalized gradient approximation (GGA) in the Perdew-Burke-Ernzerhof (PBE) form are chosen for the DFT calculations, and the projector augmented wave (PAW) is utilized to describe the electron-ion interactions. A vacuum space of 10 angstroms is inserted between periodic images to avoid interactions, and the cutoff energies of the wave function expanded by plane waves were chosen to be 500 eV. For calculations of the electronic properties and the optimal geometric structures, the first Brillouin zones are sampled by $100\times100\times1$ and $12\times12\times1$ \emph{k}-points by the Monkhorst-Pack scheme. The convergence of the Helmann-Feymann force is set to 0.01 eV ${\AA}^{-1}$.

\vskip 0.6 truecm
\par\noindent
{\bf 3. Results and discussion}
\vskip 0.3 truecm

To determine the optimal geometric structures, we calculate the graphene ripples in different longitudinal structures, ratio of height and width, and periods. Based on the structure along the periodically corrugated direction, graphene ripples come in two classical types: an armchair graphene ripple (AGR) and a zigzag graphene ripple (ZGR), as shown in Figs. 1a and 1b. The period length ($l$) is characterized by the number ($N_{\lambda}$) of zigzag lines (for AGR) or dimer lines (for ZGR) along the $y$-direction within a period. The ratio of height and width $C_{r}$, defined as the ratio of squared height to width ($h^{2}/lb$), is the main factor in determining the geometric and electronic properties, where $h$ is the amplitude of the ripple and $b$ is the C-C bond length, as shown in Fig. 1c. 3.	In our calculation, ripples with various period length l's are modelled as sinusoidal graphene ripples. We change the amplitude and frequency of sine wave to obtain different $C_{r}$'s.  When $C_{r}$ is too large, the C-C bond breaks at the crest or though of the ripple after the calculation. With an the increase of $C_{r}$, the changes in the C-C bond lengths at the crests and troughs are greater than those at the other regions. Such bond lengths are very sensitive to the longitudinal structure, where the zigzag ripple has three different nearest C-C bond lengths, while armchair ripple only has two. For $N_{\lambda}=6$ AGR and $N_{\lambda}=5$ ZGR, the largest bond length changes with the same $C_{r}$ at the crests and troughs are $1.2 \%$ and $7.9 \%$, respectively, meaning that the latter has a more seriously deformed structure. The variations of bond lengths result in different nearest-neighbor hopping integrals,\cite{kane1997size,shyu2002electronic} and thus further impacts the low-lying energy bands. The effects of curvature brings about the misorientation of the $2$p$_{z}$ orbitals and the orbital hybridization of ($2$s,$2$p$_{x}$,$2$p$_{y}$,$2$p$_{z}$), and therefore significant changes in the $\pi$ and $\sigma$ bondings can be observed. It is also noticed that the critical $C_{r}$, namely the largest ratio of height and width of a stable graphene ripple, grows with increasing $N_{\lambda}$. The highest $C_{r}$ of $N_{\lambda}=6$ AGR ($N_{\lambda}=5$ ZGR) is 0.68 (0.27), but it can reach 1.56 (1.33) in the ripple period $N_{\lambda}=12$ ($N_{\lambda}=11$). The reduction in mechanical bending force can be attributed to an increasing $C_{r}$ at a larger period. In addition to the above-mentioned curvature effects, a ripple only has a two-fold rotational symmetry along a specific direction, while a 2D graphene sheet had a six-fold rotational symmetry. These dramatic changes in the geometric symmetry are expected to play an important role in the electronic properties.

Graphene ripples exhibit very special electronic properties, being enriched by the ratio of height and width, ripple periods, and longitudinal structures. We first discuss the flat case ($C_{r}=0$) of the armchair ripple, in which the electronic states are directly sampled from those of a monolayer graphene, as shown by the black curve in Fig. 2a. Because of the chosen rectangular unit cell, energy bands of monolayer graphene are folded into a rectangular first Brillouin zone (FBZ) from its original hexagonal FBZ. Along the $\Gamma$Y direction, a pair of linearly intersecting bands cross the Fermi level ($E_{F}=0$) at $[k_{x}=0,k_{y}=2/3]$ without degeneracy. A pair of similar linear bands also makes an appearances in the direction perpendicular to $\Gamma$Y through the Fermi-momentum state (inset in Fig. 2a). This linear and isotropic behavior identifies the presence of a Dirac cone structure, and thus the intersection point is associated with the K point in the original hexagonal FBZ. These states are extracted from the energy bands of monolayer graphene that satisfy the periodic boundary condition. On the other hand, the low-lying energy bands along $\Gamma$X reveal no Dirac cone. As to the middle-energy states ($\sim \pm 2$ eV), there exists a pair of parabolic bands with a double degeneracy along $\Gamma$Y. The local minimum (maximum) of the conduction band (valence band) along $\Gamma$Y is the local maximum (minimum) along the perpendicular direction; i.e., such points are the saddle points (M and M$^{\prime}$ in the hexagonal FBZ) in the energy-wave-vector space.

The curvature causes an armchair graphene ripple to change its electronic properties dramatically at middle energy, but only subtly at low energy. The Dirac point still survives at $E_{F}=0$ and the cone structure remains isotropic, as shown by $N_{\lambda}=6$ AGR in Fig. 2a. However, the Fermi momentum shifts toward smaller $k_{y}$'s, and the Fermi velocity (the slope of the linear bands) is slightly reduced. These variations are more significant under the influence of an increasing ripple $C_{r}$, similar to those of armchair carbon nanotubes (Fig. 2a), discussed in detailed later. That the orientations of $2$p$_{z}$ orbitals are no longer parallel but instead misoriented, and that the $2$s, $2$p$_{x}$, and $2$p$_{y}$ orbitals are not in the same plane account for the change of the low-lying energy states.\cite{kane1997size} On the other side, the middle-energy parabolic bands gradually split under higher $C_{r}$, and shift towards lower energies. The change from a six-fold rotational symmetry into a two-fold one leads to the destruction of the double degeneracy of M and M$^{\prime}$. The ratio of height and width at which to start the significant splitting of the saddle point starts is $C_{r}=0.09$; it can easily be reached in experimental measurements.\cite{meng2013strain}

The low- and middle-energy states are somewhat affected by the ripple period (Fig. 2b), even when $C_{r}$ is the same. Only the isotropic Dirac cone at $E_{F}=0$ is minimally affected by a period variation. The Fermi momentum moves closer to $k_{y}=2/3$ at larger $N_{\lambda}$, and its Fermi velocity is increased, since the misorientation effect of the $2$p$_{z}$ orbitals and the hybridization of all orbitals are reduced by the larger period. The splitting of the parabolic bands of the  middle-energy saddle points can survive under different periods, meaning that the depression of rotational symmetry is non-negligible. The weaker band splitting in a larger period is attributed to the smaller changes of the C-C bond lengths. In addition, the Fermi momentum under a sufficiently large period ($N_{\lambda} \geq 22$) approaches $k_{y}=2/3$.

The longitudinal structures lead to feature-rich electronic properties. The low-energy bands of the zigzag ripple exhibit three kinds of energy dispersions: isotropic linear bands, partially flat bands, and parabolic bands, all of which exhibit a strong dependence on $C_{r}$, namely low-, middle-, and high-curvature ranges ($C_{r} \leq 0.09$, $0.09 < C_{r} < 0.19$; $C_{r} \geq 0.19$). At a low $C_{r}$, the low-lying isotropic linear bands cross with $E_{F}=0$ at the Fermi momentum $[k_{x}=3/5,k_{y}=0]$, as indicated by the $N_{\lambda}=5$ zigzag ripple in Fig. 3a. A larger $C_{r}$ is responsible for a partially flat band at about $0.4$ eV (blue curve in Fig. 3b at $C_{r}=0.10$). Closely associated with this energy dispersion are the specific carbon atoms at the ripple crests and troughs; their contributions are represented by the gray circle radius in Fig. 3b. Such band do not disappear until $C_{r}$ is at least $C_{r}=0.19$. However, two low-lying parabolic bands come into existence, as indicated in Figs. 3c and 3d at $C_{r}=0.19$ and $0.27$. They are dominated by the carbon atoms at the crests and troughs of the ripple, whose contributions are represented by the radii of the gray and black circles, respectively. The dramatic creation and destruction of energy bands are caused by a seriously deformed geometric structure, since the effects of the $2p_{z}$ misorientation and the orbital hybridization are largely enhanced. In addition, the middle-energy parabolic bands gradually split and move towards a higher energy under a larger $C_{r}$ (Fig. 3e); thus can be ascribed to the broken rotation symmetry.

The low energy band structures of ZGR are further enriched by varying the periods. When $N_{\lambda}$ grows in the manner $5\to$, $7\to$, $9\to$, $11$, the quasi-stable graphene ripples have the critical $C_{r}$'s $0.27$, $0.55$, $1.17$, $1.33$, respectively. The energy dispersions along $\Gamma$X show drastic changes, and more critical points emerge. Besides the above-mentioned low-lying parabolic bands, $N_{\lambda}=5$ ZGR exhibits a local minimum point at $E^{c} \sim 1$ eV, as shown by the black arrows in Fig. 3f. The newly formed extreme points are clearly presented by $N_{\lambda}=7$, $9$; $11$, where they have four, five, and six critical points, respectively (arrows in Figs. 3g-3i). It can be noticed that the critical points are obviously asymmetric about $E_{F}=0$, where the $N_{\lambda}=11$ ZGR shows six critical points in the conduction bands, but no such points in the valence bands. As to the $N_{\lambda}=9$ ZGR, a partially flat band comes into existence at about $0.2$ eV. A weaker energy dispersion is displayed by the $N_{\lambda}=11$ ZGR. A quasi-flat band appears at $E^{c} \sim 0.1$ eV along $\Gamma$X, while the dispersion is parabolic along $\Gamma$Y, and thus this band is a composite band. These dramatic changes about both critical points and energy dispersions mainly arise from the serious changes of the C-C bond length and strong hybridizations of the orbitals in a large $C_{r}$.

The dramatic transformation of the low-energy band structure is clearly revealed in the energy-wave-vector space. Both armchair and zigzag ripples exhibit the Dirac cone structure at a low $C_{r}$, as shown in Figs. 4a and 4c separately corresponding to $C_{r}=0.09$ for $N_{\lambda}=6$ AGR and $C_{r}=0.05$ for $N_{\lambda}=5$ ZGR. Such a structure remains isotropic for the former even at a high $C_{r}$, as displayed in Fig. 4b at $C_{r}=0.53$. However, the latter exhibits a distorted, destroyed or even shifted structure under a higher $C_{r}$. In the higher curvature range $0.09 < C_{r} < 0.19$, the conduction Dirac cones are distorted, while the valence ones remain isotropic, as indicated in Fig. 4d at $C_{r}=0.10$. Also, the energy band at about $0.4$ eV (blue curve in Fig. 3b) is partially flat in the energy-wave-vector space (Fig. 4d). At high curvatures, the Dirac point disappears at $E_{F}=0$ being either destroyed (Fig. 4e at $C_{r}=0.19$) or shifted (Fig. 4f at $C_{r}=0.27$). On the other hand, the band-edge states of the two parabolic bands in Figs. 3c and 3d that belong to the saddle points or the local maximum or minimum points are clearly displayed in Figs. 4e and 4f. These critical points strongly depend on the magnitude of $C_{r}$. In all, the transformation between two distinct band structures could be achieved by tuning the ripple curvature and the period, as seen, for example, by the dramatic transformation between the Dirac cone structure and the parabolic band (partially flat band) structure.

The variation of carrier density ($\Delta \rho$) is very useful to understand the charge bondings of all orbitals and thus the low-energy band structures. $\Delta \rho$ is created by subtracting the carrier density of an isolated carbon (hollow circle) from that of a graphene ripple as illustrated in Figs. 5a-5d. The region between the crest and the trough of the ripple, the enclosed rectangle in Figs. 5a and 5c, has a charge distribution similar to that of flat graphene. $\Delta \rho$ clearly demonstrates that carbon atoms contribute their four valence electrons to form various bonds, resulting in the reduced charge density near the atomic site (region (I)). The planar $2$s, $2$p$_{x}$, and $2$p$_{y}$ orbitals induce strong $\sigma$ bonds, leading to the increased charge density between two atoms (dumbbell shapes in (II)). Moreover, all the parallel $2p_{z}$ orbitals form $\pi$ bonds, so that there are higher charge densities normal to the surface (III). However, at crests and troughs of the ripple, the $2p_{z}$ orbitals are asymmetric about the curved surface, where $\Delta \rho$ is higher for the outer surface and lower for the inner surface, the main cause for the variation of energy bands. Such charge distributions are expected to provide an ideal chemical environment for hydrogen or lithium atoms to be adsorbed,$^{3-4}$ which shows great potential for energy storage.

The strong non-uniform curvature effect is clearly revealed by the charge distributions at the crests and troughs. The misorientation of $2p_{z}$ orbitals and strong hybridization of four orbitals, respectively, induces the extra $\sigma$ bondings and the complicated $\pi$ and $\sigma$ bondings. As for AGRs, the armchair longitudinal structure causes two distinct charge distributions along three nearest-neighbor directions, as seen in the carbon nanotube (detailed later in Fig. 5e). In other words, there exist two different bond strengths (or bond lengths), which, thus, results in two distinct nearest-neighbor hopping integrals from the point of view of the tight-binding model.\cite{kane1997size,shyu2002electronic} This is responsible for the shift of the Dirac point and the variation of the Fermi velocity. Moreover, the higher-curvature AGR (Fig. 5b) shows more significant changes of the charge distributions and the misorientation of $2p_{z}$ orbitals than a lower curvature one (Fig. 5a), as do the energy bands. On the other hand, a ZGR possesses three distinct charge distribution for three C-C bond lengths or for nearest-neighbor hopping integrals. Besides the misorientation effect at low curvature (Fig. 5c), a ZGR also displays the strong covalent bondings associated with the four orbitals at the crests and troughs when the curvature is sufficiently high ((IV) in Fig. 5d). Such strong interactions are directly reflected in the formation of the critical points of the low-energy parabolic bands (Fig. 3f), which largely enhance the density of states near $E_{F}=0$. In short, both AGRs and ZGRs exhibit a misorientation and hybridization of atomic orbitals, ZGRs in particular present these effects in a strong manner. The stronger covalent bondings at crests and troughs of ZGRs well account for the distortion, shift, and destruction of the Dirac cone band structure.

The low-lying energy bands of 2D AGRs (ZGRs) along the $\Gamma$Y are similar to those of 1D armchair (zigzag) nanotubes, further illustrating the importance of orbital hybridization and misorientation. A carbon nanotube, which satisfies a closed periodic boundary condition, has the $2p_{z}$ orbitals arranged perpendicular to the cylindrical surface but not in parallel to one another, so the curvature effect leads to small fluctuations of the hopping integrals along different directions.\cite{kane1997size,shyu2002electronic} The nearest hopping integrals of an armchair (zigzag) nanotube are given by two (three) different values, a behavior similar to that of AGRs (ZGRs). The (6,6) armchair nanotube possesses a band structure similar to that of $N_{\lambda}=6$ AGR, as shown by the dashed curve in Fig. 2a. There is a pair of linearly intersecting bands at the Fermi level and higher-energy parabolic bands. However, the Fermi momentum and Fermi velocity are distinct in armchair nanotubes and AGRs due to the uniform and non-uniform curvature effects at each atomic site. Moreover, zigzag nanotubes and ZGRs both exhibit similar low-energy bands, as indicated by the dashed curve in Fig. 3e. A small band gap appears at the Fermi momentum $[k_{y}=0]$ for a (5,0) zigzag nanotube, and a similar result is shown by the high-curvature ZGRs. The most important difference between AGRs (ZGRs) and armchair (zigzag) nanotubes in the middle-energy bands is that the former are doublely degenerate, a result directly reflected in the cylindrical symmetry.

There exist certain important differences between the zigzag and armchair ripples. A zigzag ripple exhibits three kinds of low-energy dispersions: intersecting linear bands, partially flat bands, and parabolic bands, whereas the armchair ripple only possesses the first kind. At a higher ripple curvature, the Dirac cone can be destroyed in zigzag ripples but not in armchair ones. The feature-rich energy bands are attributed to various degrees of the orbital misorientation and hybridization, which can be clearly illustrated by the charge distributions. On the other side, ARPES experiments have been successfully used to measure the band dispersions of few-layer graphenes. With the same instrument, verifications can be performed on the middle-energy band splitting induced by the reduced rotational symmetry and the two-dimensional, highly anisotropic low-energy bands caused by the curvature effect.

The density of states (DOS) directly reflects the primary characteristics of the band structures. The DOS in monolayer graphene (black curve in Fig. 6a) is zero at $E_{F}=0$, exhibits a linear $\omega$-dependence at low energy, and possesses a symmetric logarithmic divergent peak at $\omega \sim \pm 2$ eV. Such features, respectively, come from the Dirac point, intersecting linearly bands, and the saddle points. For AGRs, there are no free carrier at $E_{F}=0$ under various curvatures, owing to the hard-to-destroy Dirac point, as shown in Fig. 6a. The linear DOS is enhanced with increasing ripple curvature, a direct reflection of the reduced Fermi velocity. The split double peaks exist at middle energy, and gradually separate further for higher curvatures. On the other hand, larger periods lead to the decreased $\omega$-dependence and a smaller splitting of middle-energy peaks, as a result of the enhancement of the linear energy dispersions for the former and the weaker band splitting of the parabolic band for the latter, as indicated in Fig. 6b.

The important differences between zigzag and armchair systems lie in the low-energy DOS. ZGRs exhibit more low-lying peak structures at an increased curvature, as shown for $N_{\lambda}=5$ in Fig. 7a. The vanishing DOS at $E_{F}=0$ clearly illustrates that the ZGRs remain gapless semiconductors when $C_{r} < 0.19$. A prominent peak comes into existence at about $0.5$ eV as $C_{r}$ exceeds $0.10$ (blue curve), mainly owing to the partially flat bands. This peak will move to a lower energy with the increment of curvature. However, the DOS becomes finite at $E_{F}=0$, representing the existence of free carriers or semi-metallic characteristic under higher curvature ($C_{r} > 0.19$). That is to say, the prominent peak of the saddle point moves very close to $E_{F}=0$ (green and purple curves). Moreover, there are some weak peaks or shoulder structures at a slightly higher energy, which originate from composite bands or the extreme points. In addition, the middle-energy peaks are suppressed and gradually have a wider splitting at a larger $C_{r}$, a response associated with the parabolic bands. For a larger period, a further increase of the peak number directly reflects the more sophisticated band structures, as shown in Fig. 7b.

Both AGRs and ZGRs demonstrate certain important characteristics that are worth comparing to the experimental results. The main features of the DOSs are partially in agreement with those from the experimental measurements performed on graphene ripples using STS.\cite{meng2013strain} Two kinds of measured energy spectra are observed in the $dI/dV$-$V$ diagram, one being in a V shape, and the other exhibiting some peaks at low energy. A clear conclusion is not reached in this study, but our calculations deduce that the root cause might arise from the distinct longitudinal structures. The V-shaped diagram is associated with the DOS of AGR, which only reveals a linear $\omega$-dependence without any special structures, while the latter possesses a prominent peak near $E_{F}=0$ and some weak low-lying peaks, which are consistent with the DOS of ZGR. As to the deformation-induced quasi-Landau levels, the previous work might present a linear relationship between the peak energy and the square root of the effective quantum number. A similar linear fitting is also obtained by our calculation, as done for the $N_{\lambda}=5$ ZGR (inset in Fig. 7a). However, the large periods ($N_{\lambda}=7$, $9$ and $11$ in Fig. 7b) can induce an increase and an asymmetry for the effective quantum numbers. For example, $N_{\lambda}=11$ ZGR shows $5$ positive effective quantum numbers, while it reveals no negative ones. This asymmetry is not verified by experiment measurements nor predicted in previous tight-binding calculations.\cite{meng2013strain,guinea2008midgap} The feature-rich DOS can be further examined by STS measurements.\cite{wilder1998electronic,odom1998atomic,ouyang2001energy}

The special charge distribution of carbon atoms at the curve surface can provide a suitable environment for chemical bondings, a prerequisite for the drastic change of the electronic properties. For example, $N_{\lambda}=6$ AGR can be doped with two stripes of oxygen atoms at the crest and trough of the ripple. The oxygen atoms and the carbon atoms form strong covalent bonds so that the characteristics of $\pi$-bonds in pure carbon systems are completely destroyed. The four orbitals re-hybridize considerably, and there is no way to distinguish the contribution of $\pi$ or $\sigma$ bonds. Rippled graphene oxide displays a direct band gap at about 1.4 eV (see Fig. S1b, Supporting Information). This band gap originates mainly from the introduction of oxygen atoms, as indicated by the blue circle. The semiconducting behavior induced by surface engineering shows potential promise for applications in field-effect transistors and photocatalyst materials.

\vskip 0.6 truecm
\par\noindent
{\bf 4. Conclusion }
\vskip 0.3 truecm


In summary, the geometric and electronic properties of graphene ripples are investigated by \emph{ab initio} density functional theory calculations. Different corrugation directions, curvatures and periods are considered to exhibit different chemical and physical properties, and strongly change the optimized geometric structures, charge distributions, energy bands, and DOS. The misorientation and hybridization of carbon orbitals are associated with a variation in bond lengths and carrier densities, and are the causes for the dramatic changes in the electronic structure. The revelation of the band structures is highlighted by the two-dimensional highly anisotropic energy bands, the absence or existence of linear dispersions, newly formed critical points, and the split middle-energy states. The former three are directly induced by the special charge bondings on the crests and troughs of the ripples, which behave very differently for AGRs and ZGRs, while the last one originates from the reduced rotational symmetry. Moreover, the similarity between graphene ripples and carbon nanotubes  can be explained by the curvature effect. The predicted band dispersions could be further measured by ARPES. As for DOS, the semimetal-semiconductor transitions, the increasing number of low-lying peaks, and the splitting middle-energy peaks are, respectively, owed to the highly anisotropic saddle structure near $E_{F}=0$, the appearance of local minimum or maximum points, and the split parabolic bands. These more sophisticated features can be further tested by the STS measurements. Our results suggest a route to distinguish specific types of graphene ripples obtained in STS experiments, and could be further reinforced by future ARPES experiments. The corrugated ripple structures are very suitable for chemical doping and engineering, and the tunable electronic properties might be potentially important for applications in nanoelectronic devices and energy storage.

\par\noindent {\bf Acknowledgments}

This work is supported by the NSC and NCTS (South) of Taiwan, under the grant Nos. NSC-102-2112-M-006-007-MY3.

\newpage
\renewcommand{\baselinestretch}{0.2}

\begin{figure}[htbp]
\center
\rotatebox{0} {\includegraphics[width=14cm]{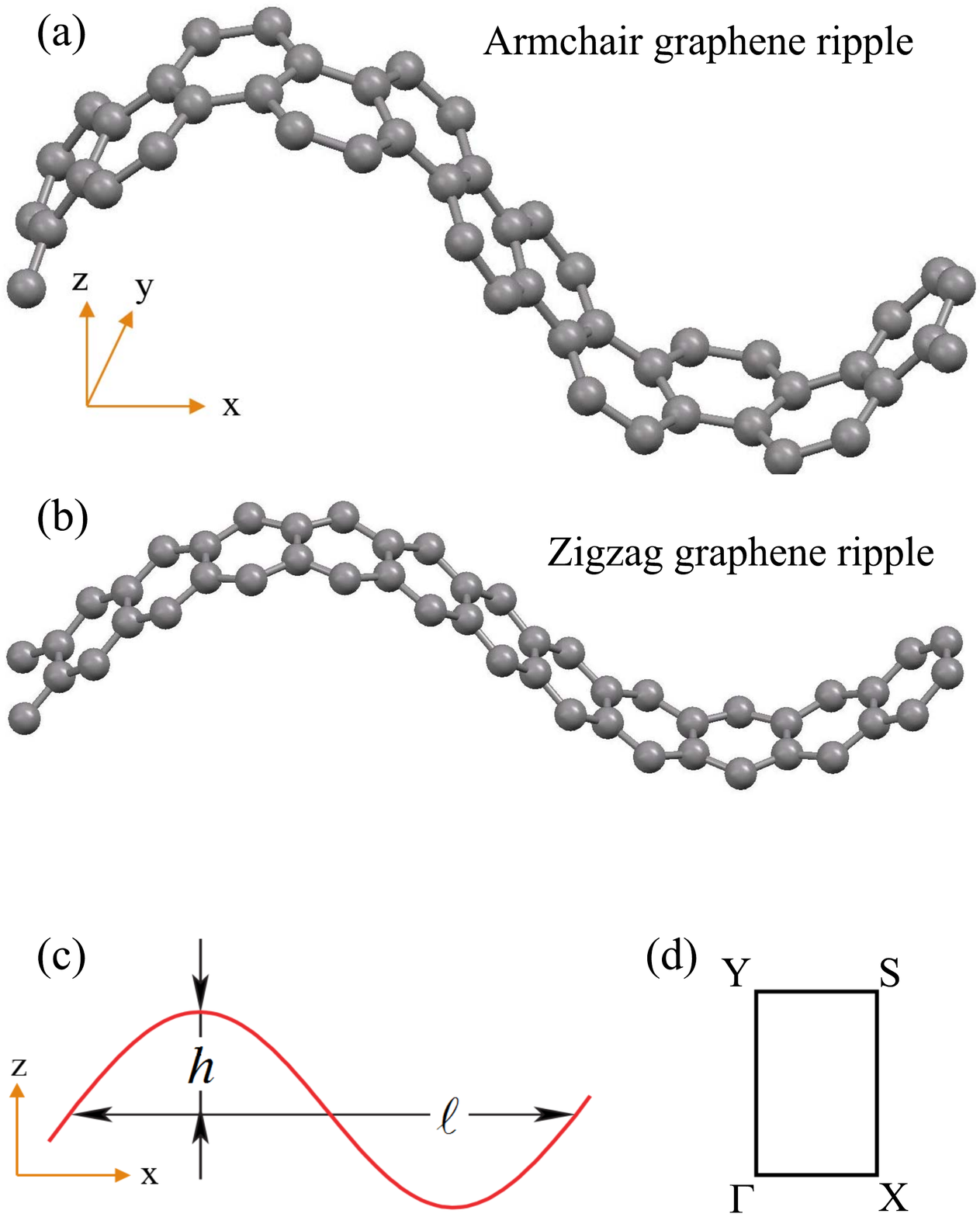}}
\caption{Geometric structures of (a) armchair and (b) zigzag ripples. (c) Side view of the graphene ripple with the rippling period denoted by $l$ and the amplitude by $h$. (d) The rectangular first Brillouin zone of graphene ripple.}
\label{Figure 1}
\end{figure}

\begin{figure}[htbp]
\center
\rotatebox{0} {\includegraphics[width=14cm]{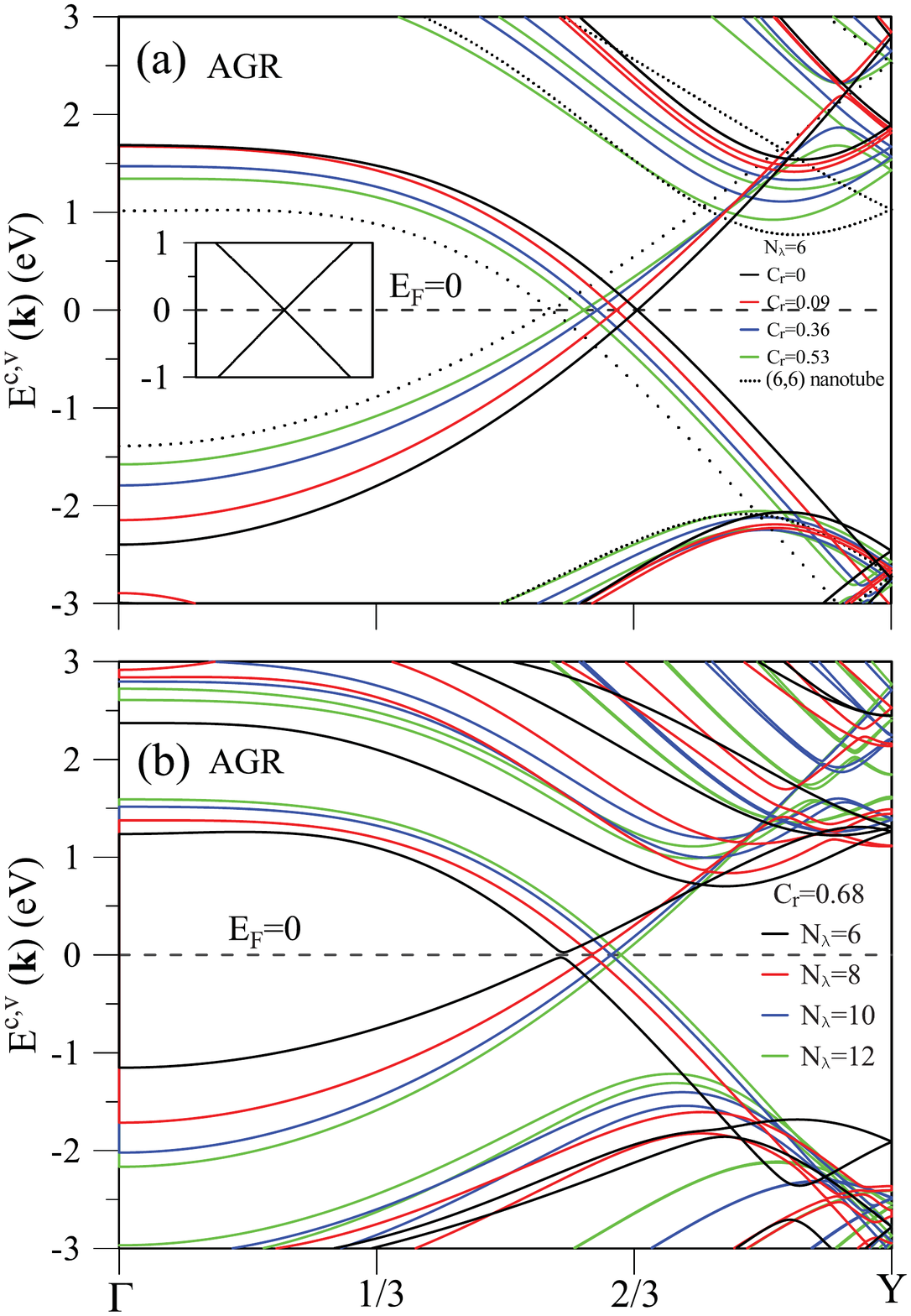}}
\caption{Energy bands of armchair graphene ripples with various (a) ripple curvatures and (b) periods.}
\label{Figure 2}
\end{figure}

\begin{figure}[htbp]
\center
\rotatebox{0} {\includegraphics[width=14cm]{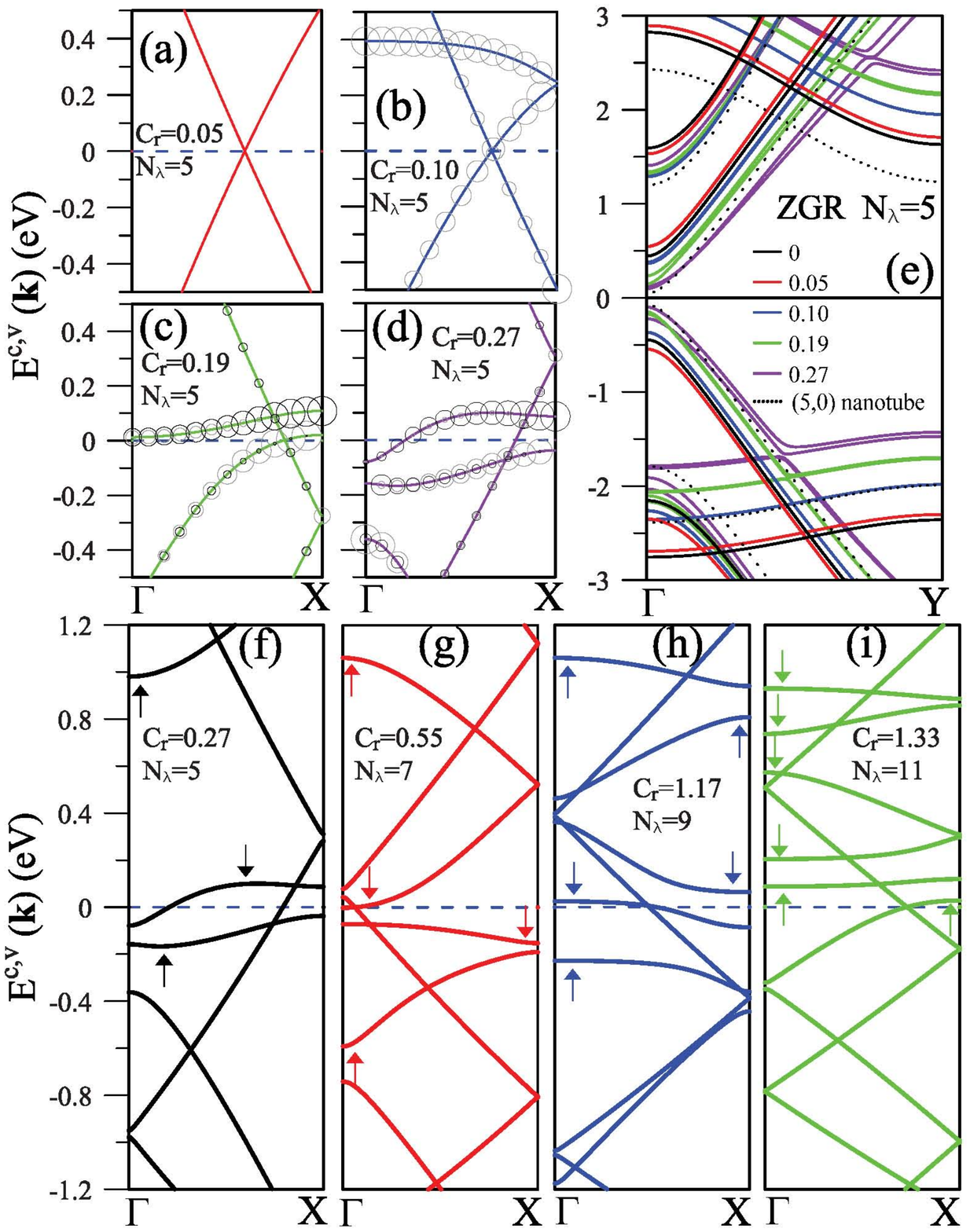}}
\caption{Energy bands of a zigzag graphene ripple along the  (a)-(d) $\Gamma$X and (e) $\Gamma$Y direction. (f)-(i) Energy bands for the critical ripple curvature with increasing ripple periods.}
\label{Figure 3}
\end{figure}

\begin{figure}[htbp]
\center
\rotatebox{0} {\includegraphics[width=14cm]{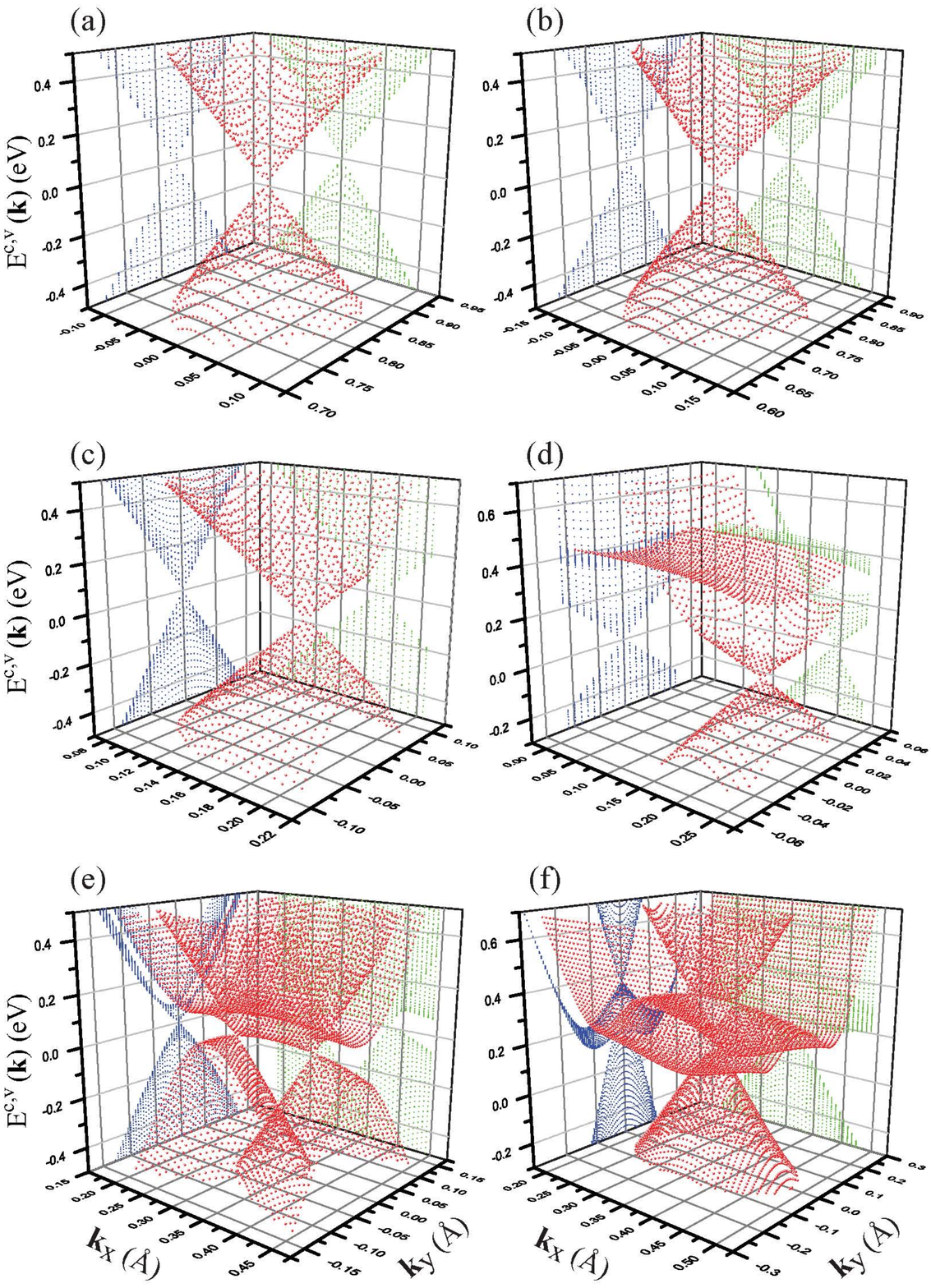}}
\caption{Energy bands in energy-wave-vector space for armchair graphene ripples at (a) $C_{r}=0.09$ and (b) $C_{r}=0.53$, and zigzag ripples at (c) $C_{r}=0.05$, (d) $C_{r}=0.10$, (e) $C_{r}=0.19$, and (f) $C_{r}=0.27$.}
\label{Figure 4}
\end{figure}

\begin{figure}[htbp]
\center
\rotatebox{0} {\includegraphics[width=14cm]{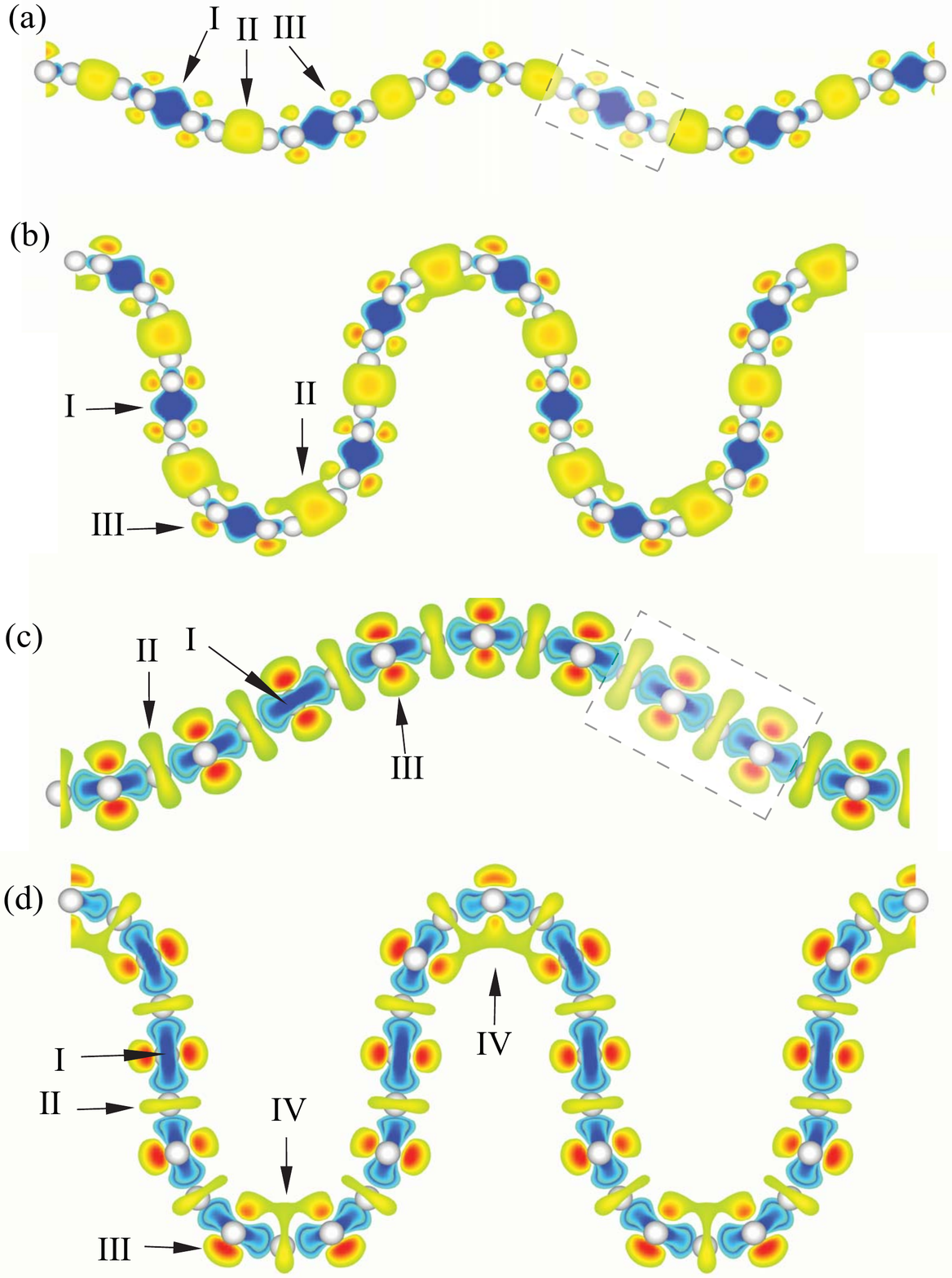}}
\caption{Charge distribution of armchair graphene ripples at (a) low and (b) high curvatures, and zigzag graphene ripples at (c) low and (d) high curvatures.}
\label{Figure 5}
\end{figure}

\begin{figure}[htbp]
\center
\rotatebox{0} {\includegraphics[width=14cm]{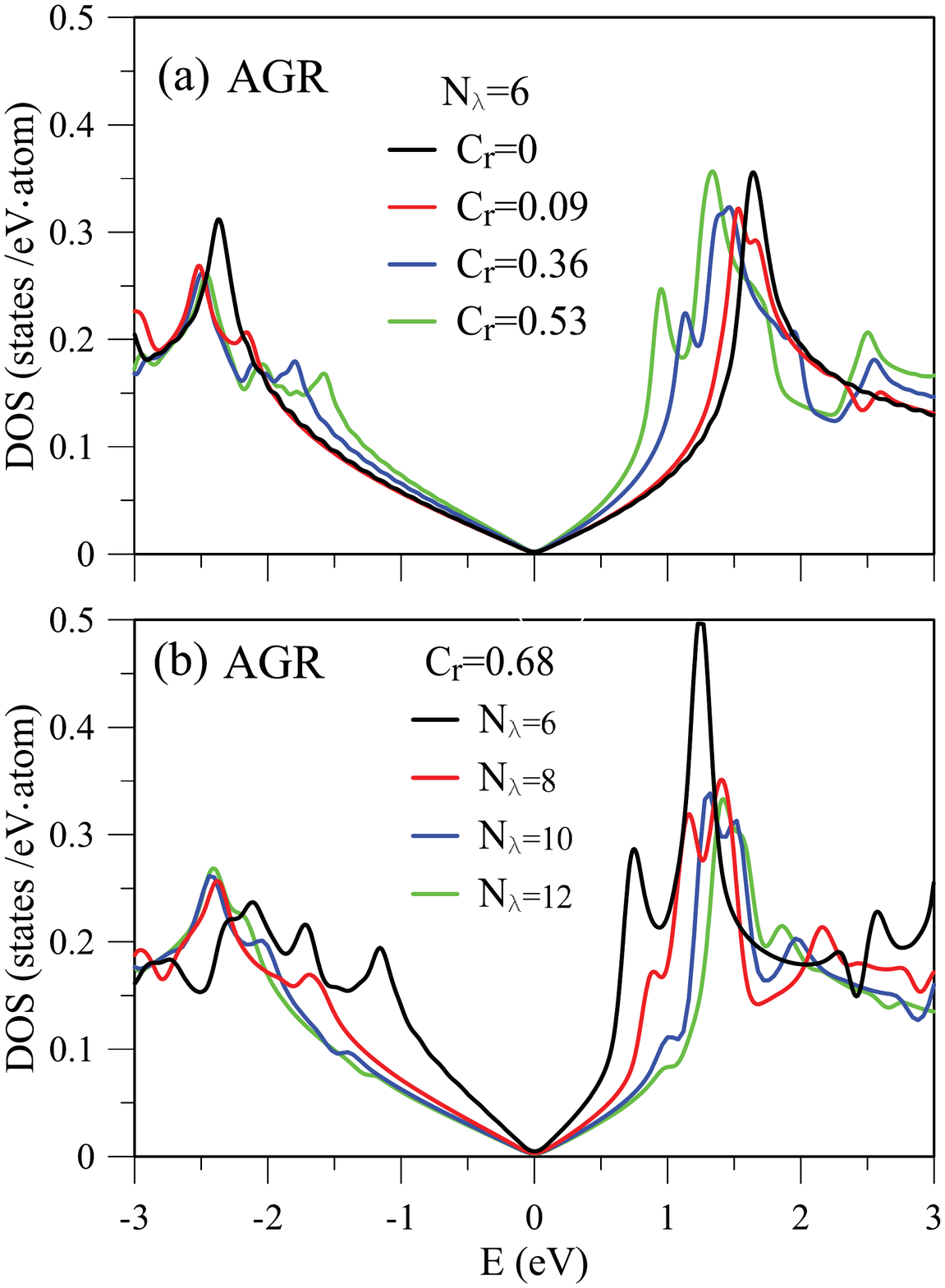}}
\caption{Density of states of armchair graphene ripples with various (a) ripple curvatures and (b) periods.}
\label{Figure 6}
\end{figure}

\begin{figure}[htbp]
\center
\rotatebox{0} {\includegraphics[width=14cm]{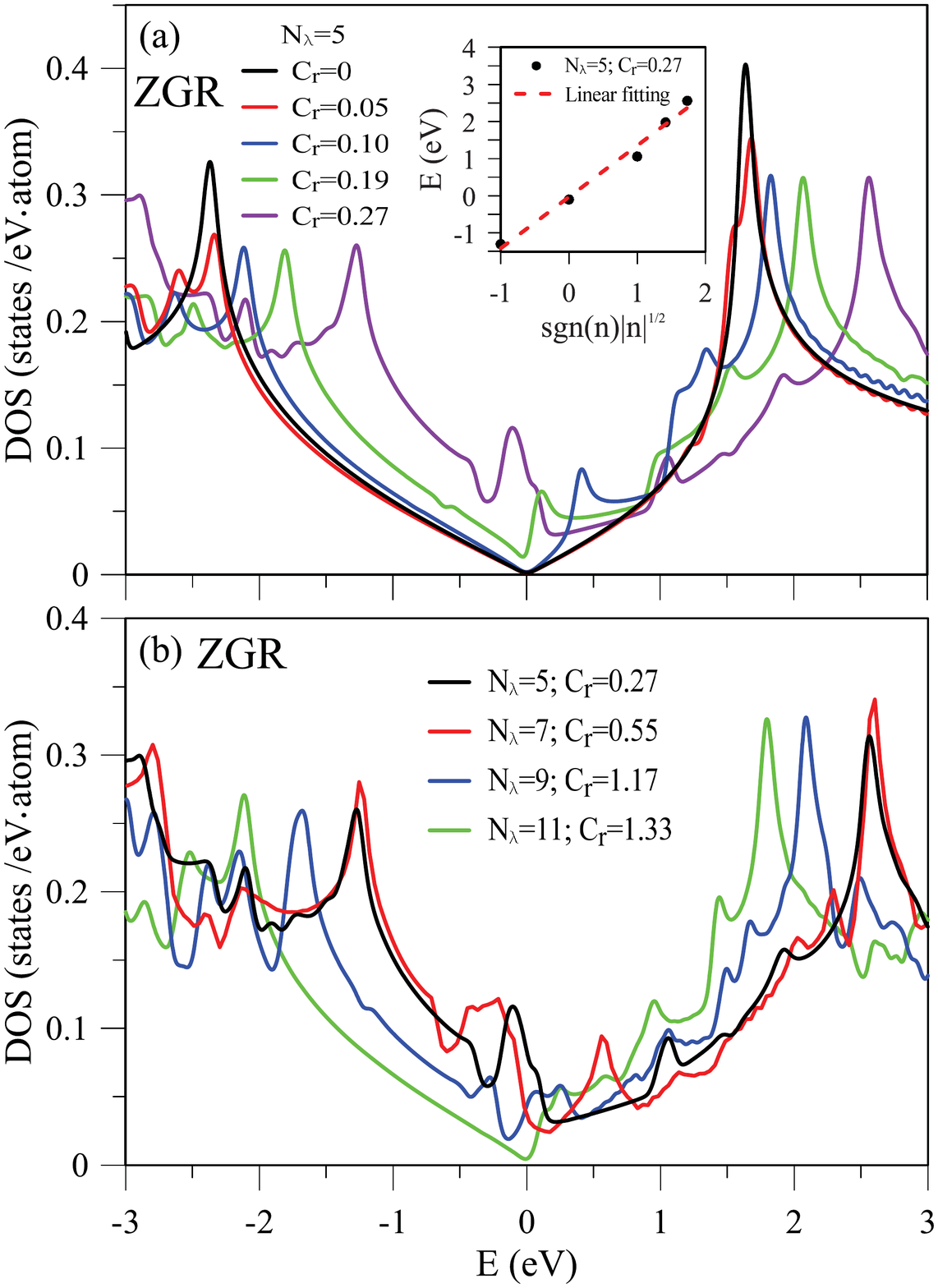}}
\caption{Density of states of zigzag graphene ripples with various (a) ripple curvatures and (b) periods.}
\label{Figure 7}
\end{figure}


\begin{thebibliography}{99}
\bibitem{novoselov2005two} Novoselov KS, Geim AK, Morozov SV, Jiang D, Grigorieva MKI, Dubonos SV, et al. Two-dimensional gas of massless Dirac fermions in graphene. Nature 2005, 438:197-200.

\bibitem{geim2009graphene} Geim AK. Graphene: status and prospects. Science 2009, 324(5934):1530-1534.

\bibitem{lee2008measurement} Lee C, Wei X, Kysar JW, Hone J. Measurement of the elastic properties and intrinsic strength of monolayer graphene. Science 2008, 321(5887):385-388.

\bibitem{stankovich2007synthesis} Stankovich S, Dikin DA, Piner RD, Kohlhaas KA, Kleinhammes A, Jia Y, et al. Synthesis of graphene-based nanosheets via chemical reduction of exfoliated graphite oxide. Carbon 2007, 45(7):1558-1565.

\bibitem{soldano2010production} Soldano C, Mahmood A, Dujardin E. Production, properties and potential of graphene. Carbon 2010, 48(8):2127-2150.

\bibitem{sprinkle2009first} Sprinkle M, Siegel D, Hu Y, Hicks J, Tejeda A, Taleb-Ibrahimi A, et al. First direct observation of a nearly ideal graphene band structure. Physical Review Letters 2009, 103(22):226803.

\bibitem{hwang2007carrier} Hwang EH, Adam S, Sarma SD. Carrier transport in two-dimensional graphene layers. Physical Review Letters 2007, 98(18):186806.

\bibitem{sarma2011electronic} Sarma SD, Adam S, Hwang EH, Rossi E. Electronic transport in two-dimensional graphene. Reviews of Modern Physics 2011, 83(2):407.

\bibitem{zhang2006landau} Zhang Y, Jiang Z, Small JP, Purewal MS, Tan YW, Fazlollahi, et al. Landau-level splitting in graphene in high magnetic fields. Physical Review Letters 2006, 96(13):136806.

\bibitem{checkelsky2008zero} Checkelsky JG, Li L, Ong NP. Zero-energy state in graphene in a high magnetic field. Physical Review Letters 2008, 100(20):206801.

\bibitem{bolotin2008ultrahigh} Bolotin KI, Sikes KJ, Jiang Z, Klima M, Fudenberg G, Hone J, et al. Ultrahigh electron mobility in suspended graphene. Solid State Communications 2008, 146(9):351-355.

\bibitem{wang2009high} Wang S, Ang PK, Wang Z, Tang ALL, Thong JT, Loh KP. High mobility, printable, and solution-processed graphene electronics. Nano Letters 2009, 10(1):92-98.

\bibitem{zhang2005experimental} Zhang Y, Tan YW, Stormer HL, Kim P. Experimental observation of the quantum Hall effect and Berry's phase in graphene. Nature 2005, 438(7065):201-204.

\bibitem{gusynin2005unconventional} Gusynin VP, Sharapov SG. Unconventional integer quantum Hall effect in graphene. Physical Review Letters 2005, 95(14):146801.

\bibitem{du2009fractional} Du X, Skachko I, Duerr F, Luican A, Andrei EY. Fractional quantum Hall effect and insulating phase of Dirac electrons in graphene. Nature 2009, 462(7270):19.

\bibitem{cortijo2007effects} Cortijo A, Vozmediano MA. Effects of topological defects and local curvature on the electronic properties of planar graphene. Nuclear Physics B 2007, 763(3):293-308.

\bibitem{lin2014magneto} Lin CY, Wu JY, Chang CP, Lin MF. Magneto-optical selection rules of curved graphene nanoribbons and carbon nanotubes. Carbon 2014, 69:151-161.

\bibitem{chang2014geometric} Chang SL, Lin SY, Lin SK, Lee CH, Lin MF. Geometric and Electronic Properties of Edge-decorated Graphene Nanoribbons. Scientific Reports 2014, 4:6038.

\bibitem{nilsson2006electronic} Nilsson J, Neto AC, Guinea F, Peres NMR. Electronic properties of graphene multilayers. Physical Review Letters 2006, 97(26):266801.

\bibitem{avetisyan2010stacking} Avetisyan AA, Partoens B, Peeters FM. Stacking order dependent electric field tuning of the band gap in graphene multilayers. Physical Review B 2010, 81(11):115432.

\bibitem{ni2008uniaxial} Ni ZH, Yu T, Lu YH, Wang YY, Feng YP, Shen ZX. Uniaxial strain on graphene: Raman spectroscopy study and band-gap opening. ACS Nano 2008, 2(11):2301-2305.

\bibitem{wong2012strain} Wong JH, Wu BR, Lin MF. Strain effect on the electronic properties of single layer and bilayer graphene. The Journal of Physical Chemistry C 2012, 116(14):8271-8277.

\bibitem{orlita2009graphite} Orlita M, Faugeras C, Schneider JM, Martinez G, Maude DK, Potemski M. Graphite from the viewpoint of Landau level spectroscopy: An effective graphene bilayer and monolayer. Physical Review Letters 2009, 102(16):166401.

\bibitem{chuang2013electric} Chuang YC, Wu JY, Lin MF. Electric field dependence of excitation spectra in AB-stacked bilayer graphene. Scientific Reports 2013, 3:1368.

\bibitem{wu2014combined} Wu JY, Gumbs G, Lin MF. Combined effect of stacking and magnetic field on plasmon excitations in bilayer graphene. Physical Review B 2014, 89(16):165407.

\bibitem{fasolino2007intrinsic} Fasolino A, Los JH, Katsnelson MI. Intrinsic ripples in graphene. Nature materials 2007, 6(11):858-861.

\bibitem{meyer2007structure} Meyer JC, Geim AK, Katsnelson MI, Novoselov KS, Booth TJ, Roth S. The structure of suspended graphene sheets. Nature 2007, 446(7131):60-63.

\bibitem{meng2013strain} Meng L, He WY, Zheng H, Liu M, Yan H, Yan W, et al. Strain-induced one-dimensional Landau level quantization in corrugated graphene. Physical Review B 2013, 87(20):205405.

\bibitem{yan2013strain} Yan W, He WY, Chu ZD, Liu M, Meng L, Dou RF, et al. Strain and curvature induced evolution of electronic band structures in twisted graphene bilayer. Nature communications 2013, 4:2159.

\bibitem{capasso2014graphene} Capasso A, Placidi E, Zhan HF, Perfetto E, Bell JM, Gu Y, et al. Graphene ripples generated by grain boundaries in highly ordered pyrolytic graphite. Carbon 2014, 68:330-336.

\bibitem{de2008periodically} De Parga AV, Calleja F, Borca BMCG, Passeggi Jr MCG, Hinarejos JJ, Guinea F, et al. Periodically rippled graphene: growth and spatially resolved electronic structure. Physical Review Letters 2008, 100(5):056807.

\bibitem{bao2009controlled} Bao W, Miao F, Chen Z, Zhang H, Jang W, Dames C, et al. Controlled ripple texturing of suspended graphene and ultrathin graphite membranes. Nature nanotechnology 2009, 4(9):562-566.

\bibitem{georgiou2011graphene} Georgiou T, Britnell L, Blake P, Gorbachev RV, Gholinia A, Geim AK, et al. Graphene bubbles with controllable curvature. Applied Physics Letters 2011, 99(9):093103.

\bibitem{xie2009controlled} Xie X, Ju L, Feng X, Sun Y, Zhou R, Liu K, et al. Controlled fabrication of high-quality carbon nanoscrolls from monolayer graphene. Nano Letters 2009, 9(7):2565-2570.

\bibitem{xia2010fabrication} Xia D, Xue Q, Xie J, Chen H, Lv C, Besenbacher F, et al. Fabrication of carbon nanoscrolls from monolayer graphene. Small 2010, 6(18):2010¡V2019.

\bibitem{liu2012folded} Liu F, Song S, Xue D, Zhang H. Folded structured graphene paper for high performance electrode materials. Advanced Materials 2012, 24(8):1089-1094.

\bibitem{kim2011multiply} Kim K, Lee Z, Malone BD, Chan KT, Aleman B, Regan W, et al. Multiply folded graphene. Physical Review B 2011, 83(24):245433.

\bibitem{yoo2008large} Yoo E, Kim J, Hosono E, Zhou HS, Kudo T, Honma I. Large reversible Li storage of graphene nanosheet families for use in rechargeable lithium ion batteries. Nano Letters 2008, 8(8):2277-2282.

\bibitem{yan2012high} Yan J, Liu J, Fan Z, Wei T, Zhang L. High-performance supercapacitor electrodes based on highly corrugated graphene sheets. Carbon 2012, 50(6):2179-2188.

\bibitem{tozzini2011reversible} Tozzini V, Pellegrini V. Reversible hydrogen storage by controlled buckling of graphene layers. The Journal of Physical Chemistry C 2011, 115(51):25523-25528.

\bibitem{hsiao2010preparation} Hsiao MC, Liao SH, Yen MY, Liu PI, Pu NW, Wang CA, et al. Preparation of covalently functionalized graphene using residual oxygen-containing functional groups. ACS applied materials and interfaces 2010, 2(11):3092-3099.

\bibitem{guinea2008midgap} Guinea F, Katsnelson MI, Vozmediano MAH. Midgap states and charge inhomogeneities in corrugated graphene. Physical Review B 2008, 77(7):075422.

\bibitem{guinea2008gauge} Guinea F, Horovitz B, Le Doussal P. Gauge field induced by ripples in graphene. Physical Review B 2008, 77(20):205421.

\bibitem{guinea2009gauge} Guinea F, Horovitz B, Le Doussal P. Gauge fields, ripples and wrinkles in graphene layers. Solid State Communications 2009, 149(27):1140-1143.

\bibitem{vozmediano2010gauge} Vozmediano MA, Katsnelson MI, Guinea F. Gauge fields in graphene. Physics Reports 2010, 496(4):109-148.

\bibitem{wehling2008midgap} Wehling TO, Balatsky AV, Tsvelik AM, Katsnelson MI, Lichtenstein AI. Midgap states in corrugated graphene: Ab initio calculations and effective field theory. Europhysics Letters 2008, 84(1):17003.

\bibitem{ohta2007interlayer} Ohta T, Bostwick A, McChesney JL, Seyller T, Horn K, Rotenberg E. Interlayer interaction and electronic screening in multilayer graphene investigated with angle-resolved photoemission spectroscopy. Physical Review Letters 2007, 98(20):206802.

\bibitem{ohta2006controlling} Ohta T, Bostwick A, Seyller T, Horn K, Rotenberg E. Controlling the electronic structure of bilayer graphene. Science 2006, 313(5789):951-954.

\bibitem{kresse1996efficient} Kresse G, Furthm\"{u}ller J. Efficient iterative schemes for ab initio total-energy calculations using a plane-wave basis set. Physical Review B 1996, 54(16):11169.

\bibitem{kane1997size} Kane CL, Mele EJ. Size, shape, and low energy electronic structure of carbon nanotubes. Physical Review Letters 1997, 78(10):1932.

\bibitem{shyu2002electronic} Shyu FL, Lin MF. Electronic and optical properties of narrow-gap carbon nanotubes. Journal of the Physical Society of Japan 2002, 71(8):1820-1823.

\bibitem{wilder1998electronic} Wilder JW, Venema LC, Rinzler AG, Smalley RE, Dekker C. Electronic structure of atomically resolved carbon nanotubes. Nature 1998, 391(6662):59-62.

\bibitem{odom1998atomic} Odom TW, Huang JL, Kim P, Lieber CM. Atomic structure and electronic properties of single-walled carbon nanotubes. Nature 1998, 391(6662):62-64.

\bibitem{ouyang2001energy} Ouyang M, Huang JL, Cheung CL, Lieber CM. Energy gaps in" metallic" single-walled carbon nanotubes. Science 2001, 292(5517):702-705.

\end{thebibliography}
\end{document}